\documentclass[twocolumn]{aastex631}

\usepackage{hyperref}

\usepackage{multirow}
\usepackage{booktabs}

\shortauthors{Giacalone et al.}

\begin{document}

\title{HD 56414 b: A Warm Neptune Transiting an A-type Star}


\correspondingauthor{Steven Giacalone}
\email{steven$\_$giacalone@berkeley.edu}

\author[0000-0002-8965-3969]{Steven Giacalone}
\affil{Department of Astronomy, University of California Berkeley, Berkeley, CA 94720, USA}

\author[0000-0001-8189-0233]{Courtney D. Dressing}
\affiliation{Department of Astronomy, University of California Berkeley, Berkeley, CA 94720, USA}

\author[0000-0003-1756-4825]{A. Garc\'ia Mu\~noz}
\affiliation{Universit\'e Paris-Saclay, Universit\'e Paris Cit\'e, CEA, CNRS, AIM, 91191, Gif-sur-Yvette, France}

\author[0000-0003-0030-332X]{Matthew~J.~Hooton}
\affiliation{Cavendish Laboratory, University of Cambridge, JJ Thomson Avenue, Cambridge, CB3 0HE, UK}

\author[0000-0002-3481-9052]{Keivan G. Stassun}
\affiliation{Department of Physics and Astronomy, Vanderbilt University, Nashville, TN 37235, USA}

\author[0000-0002-8964-8377]{Samuel N. Quinn}
\affiliation{Center for Astrophysics \textbar{} Harvard \& Smithsonian, 60 Garden Street, Cambridge, MA 02138, USA}

\author[0000-0002-4891-3517]{George Zhou}
\affiliation{University of Southern Queensland, Centre for Astrophysics, West Street, Toowoomba, QLD 4350 Australia}

\author[0000-0002-0619-7639]{Carl Ziegler}
\affiliation{Department of Physics, Engineering and Astronomy, Stephen F. Austin State University, 1936 North St, Nacogdoches, TX 75962, USA}


\author[0000-0001-6763-6562]{Roland~Vanderspek}
\affiliation{Department of Physics and Kavli Institute for Astrophysics and Space Research, Massachusetts Institute of Technology, Cambridge, MA 02139, USA}

\author[0000-0001-9911-7388]{David~W.~Latham}
\affiliation{Center for Astrophysics \textbar{} Harvard \& Smithsonian, 60 Garden Street, Cambridge, MA 02138, USA}

\author[0000-0002-6892-6948]{S.~Seager}
\affiliation{Department of Physics and Kavli Institute for Astrophysics and Space Research, Massachusetts Institute of Technology, Cambridge, MA 02139, USA}
\affiliation{Department of Earth, Atmospheric and Planetary Sciences, Massachusetts Institute of Technology, Cambridge, MA 02139, USA}
\affiliation{Department of Aeronautics and Astronautics, MIT, 77 Massachusetts Avenue, Cambridge, MA 02139, USA}

\author[0000-0002-4265-047X]{Joshua~N.~Winn}
\affiliation{Department of Astrophysical Sciences, Princeton University, 4 Ivy Lane, Princeton, NJ 08544, USA}

\author[0000-0002-4715-9460]{Jon~M.~Jenkins}
\affiliation{NASA Ames Research Center, Moffett Field, CA 94035, USA}

\author[0000-0001-7124-4094]{C\'{e}sar Brice\~{n}o}
\affiliation{SOAR Telescope/NSF’s NOIRLab, Casilla 603, La Serena, Chile} 


\author[0000-0003-0918-7484]{Chelsea~ X.~Huang}
\affiliation{University of Southern Queensland, Centre for Astrophysics, West Street, Toowoomba, QLD 4350 Australia}
\affiliation{Juan Carlos Torres Fellow}

\author[0000-0003-1286-5231]{David~R.~Rodriguez}
\affiliation{Space Telescope Science Institute, 3700 San Martin Drive, Baltimore, MD, 21218, USA}

\author[0000-0002-1836-3120]{Avi~Shporer}
\affiliation{Department of Physics and Kavli Institute for Astrophysics and Space Research, Massachusetts Institute of Technology, Cambridge, MA 02139, USA}


\author[0000-0003-3654-1602]{Andrew W. Mann}
\affiliation{Department of Physics and Astronomy, The University of North Carolina at Chapel Hill, Chapel Hill, NC 27599-3255, USA}

\author{David Watanabe}
\affiliation{Planetary Discoveries in Fredericksburg, VA 22405, USA}

\author[0000-0002-5402-9613]{Bill~Wohler}
\affiliation{NASA Ames Research Center, Moffett Field, CA 94035, USA}
\affiliation{SETI Institute, Mountain View, CA 94043, USA}
 
\published{in the Astrophysical Journal Letters on August 12, 2022}
 
\begin{abstract}
We report the discovery in TESS data and validation of HD 56414 b (a.k.a. TOI-1228 b), a Neptune-size ($R_{\rm p} = 3.71 \pm 0.20\, R_\oplus$) planet with a 29-day orbital period transiting a young (Age = $420 \pm 140$ Myr) A-type star in the TESS southern continuous viewing zone. HD 56414 is one of the hottest stars ($T_{\rm eff} = 8500 \pm 150 \, {\rm K}$) to host a known sub-Jovian planet. HD 56414 b lies on the boundary of the hot Neptune desert in planet radius -- bolometric insolation flux space, suggesting that the planet may be experiencing mass loss. To explore this, we apply a photoevaporation model that incorporates the high near ultraviolet continuum emission of A-type stars. We find that the planet can retain most of its atmosphere over the typical 1-Gyr main sequence lifetime of an A-type star if its mass is $\ge 8 \, M_\oplus$. Our model also predicts that close-in Neptune-size planets with masses $< 14 \, M_\oplus$ are susceptible to total atmospheric stripping over 1 Gyr, hinting that the hot Neptune desert, which has been previously observed around FGKM-type stars, likely extends to A-type stars.
\end{abstract}

\section{Introduction}

Over $99 \%$ of planets have been discovered around FGKM dwarfs, main-sequence stars with effective temperatures under $\sim 7500$ K \citep{habets1981}. Consequently, relatively little is known about planets orbiting hotter and more massive stars, which differ from cooler stars in multiple ways that could affect orbiting planets. For instance, A-type stars are often fully radiative, which could impede their abilities to dissipate tides excited by close-in planets, leading to slower orbital evolution due to tidal interactions relative to planets orbiting cooler stars with convective layers \citep[e.g.,][]{winn2010}. Another consequence of being fully radiative is reduced X-ray and extreme ultraviolet (XUV) emission, believed to be the main driver of atmospheric mass loss for close-in planets around FGKM-type stars \citep{lammer2003}, which suggests that planets around these stars are less likely to lose their atmospheres. Conversely, it has also been predicted that, due to their high effective temperatures and luminosities, A-type stars can drive atmospheric mass loss of close-in planets even more rapidly than cooler stars \citep[e.g.,][]{munoz2019}. Lastly, because A-type stars have shorter main sequence lifetimes than cooler stars, their planets tend to be younger, and perhaps less evolved.

In general, planets are more difficult to detect around A-type stars than around cooler stars. First, a planet of a given size orbiting a larger star produces a shallower transit. Second, rapid rotational speeds and pulsation-driven jitter complicate mass measurements of planets orbiting A stars \citep[e.g.,][]{griffin2000, galland2005}. In addition, A-type stars are intrinsically rarer than cooler stars \citep[by a factor of $\sim 100$;][]{vanleeuwen2007}, which reduces the sample around which we can search for these planets.

Nonetheless, several studies have constrained the occurrence rate of giant planets around these hot stars. \cite{johnson2010} calculated the frequency of planets more massive than Jupiter with orbital separations $<2.5$ AU by collecting radial velocities (RVs) of main sequence FGKM stars as well as somewhat more massive stars stars that may have evolved off the main sequence (and consequently have slower rotation rates). They found that giant planets generally become more common as stellar mass increases. \cite{zhou2019} measured the occurrence rate of transiting hot Jupiters around A-type stars and found no statistically significant difference compared to FG-type stars. Lastly, \cite{nielsen2019} measured the occurrence of directly imaged giant planets with orbital separations between 3 and 100 AU and also found a strong correlation between occurrence rate and stellar mass.

While these studies provided valuable insight into the population of giant planets around A-type stars, the population of smaller and less massive planets is largely unexplored: only three planets smaller than Jupiter (hereafter referred to as ``sub-Jovian'') have been confirmed around these hot stars: Kepler-1115 b \citep{morton2016}, Kepler-462 b \citep{morton2016}, and Kepler-462 c \citep{masuda2020}. A larger sample is required to investigate the properties and frequency of the smaller and less massive planets orbiting A-type stars. In this paper, we add another planet to this sample: the warm Neptune HD 56414 b.

This paper is structured as follows. In Section \ref{sec: observations}, we present the observations used in our analysis. In Section \ref{sec: analysis} we characterize the planet and star. In Section \ref{sec: validation} we validate the planet. Lastly, in Section \ref{sec: discussion}, we address the significance of the newly discovered planet and conclude.

\begin{figure*}[h!]
    \centering
    \includegraphics[width=1.0\textwidth]{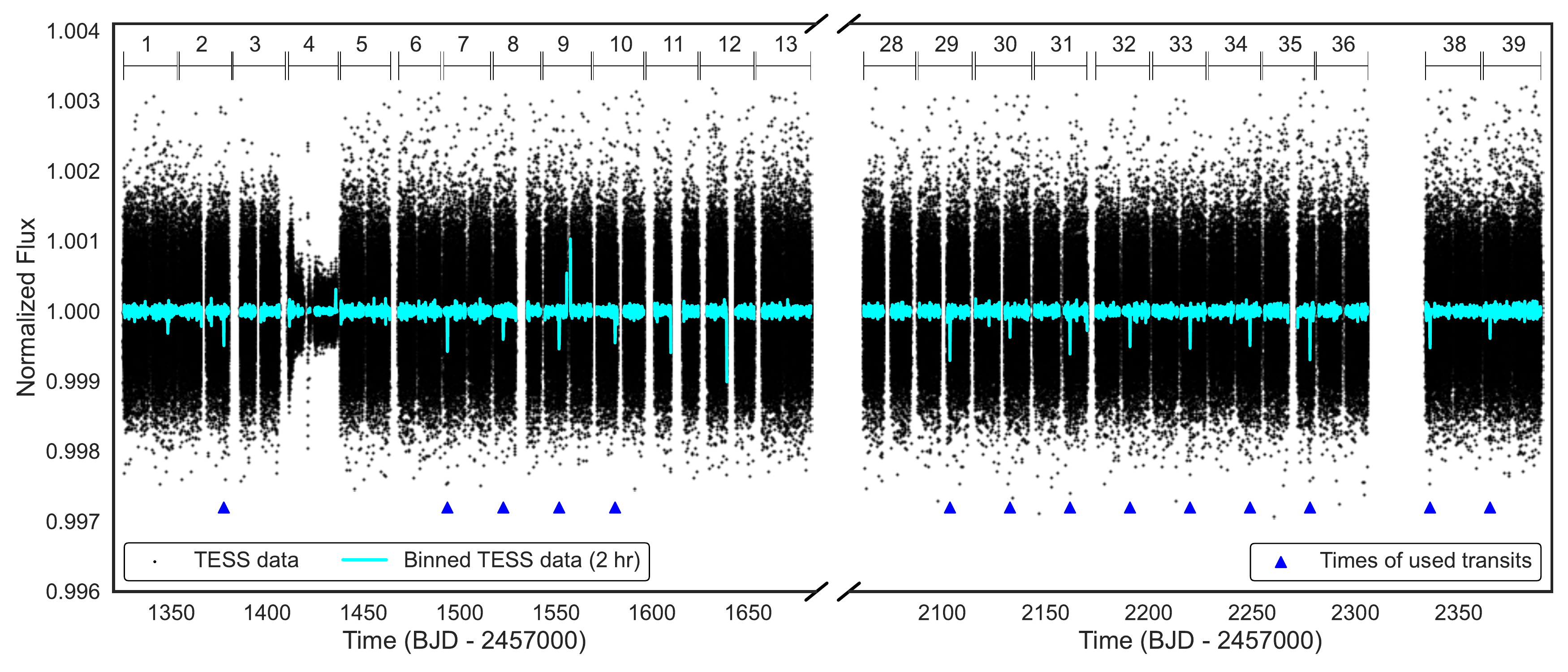}
    \includegraphics[width=0.49\textwidth]{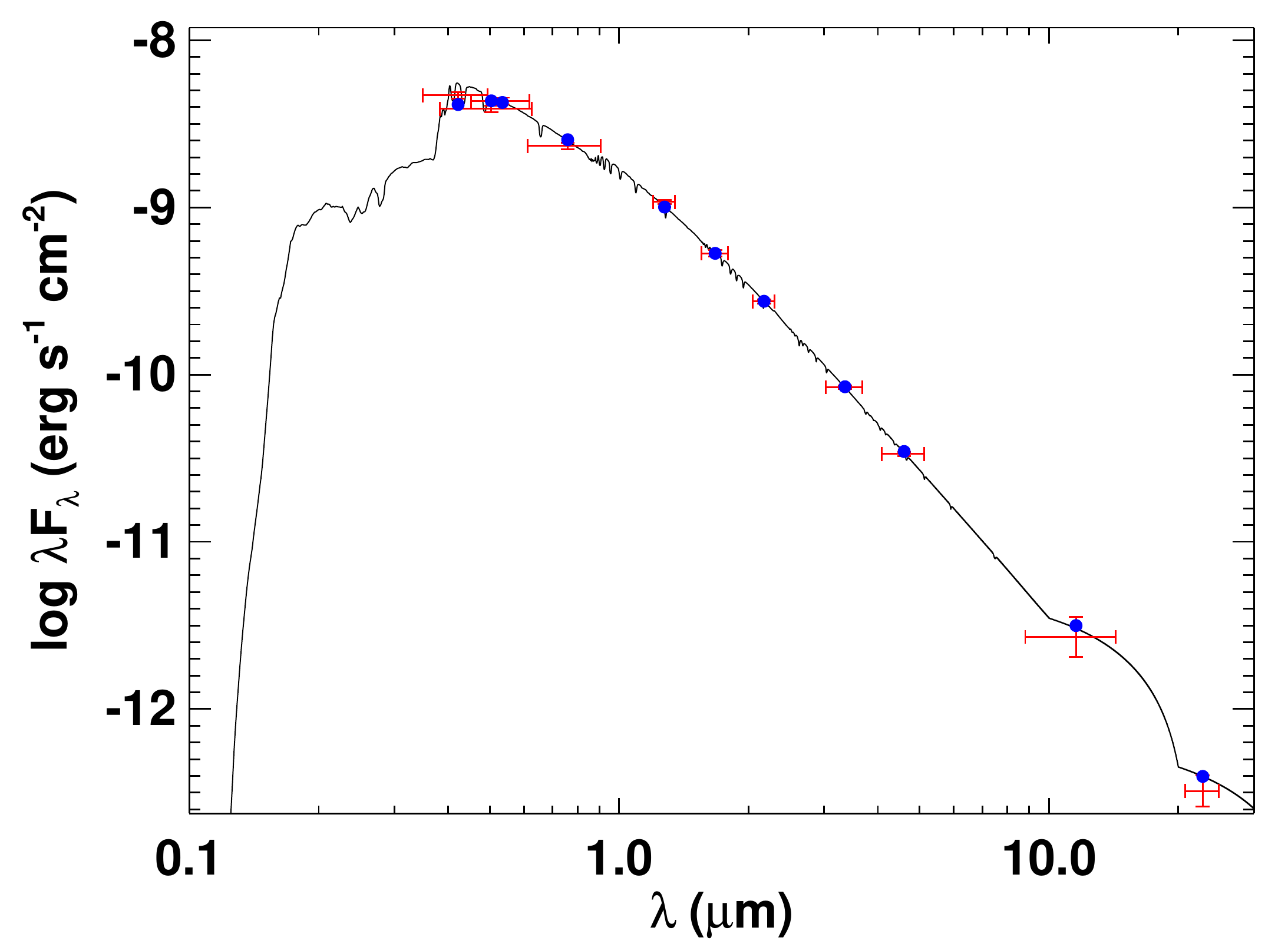}
    \includegraphics[width=0.49\textwidth]{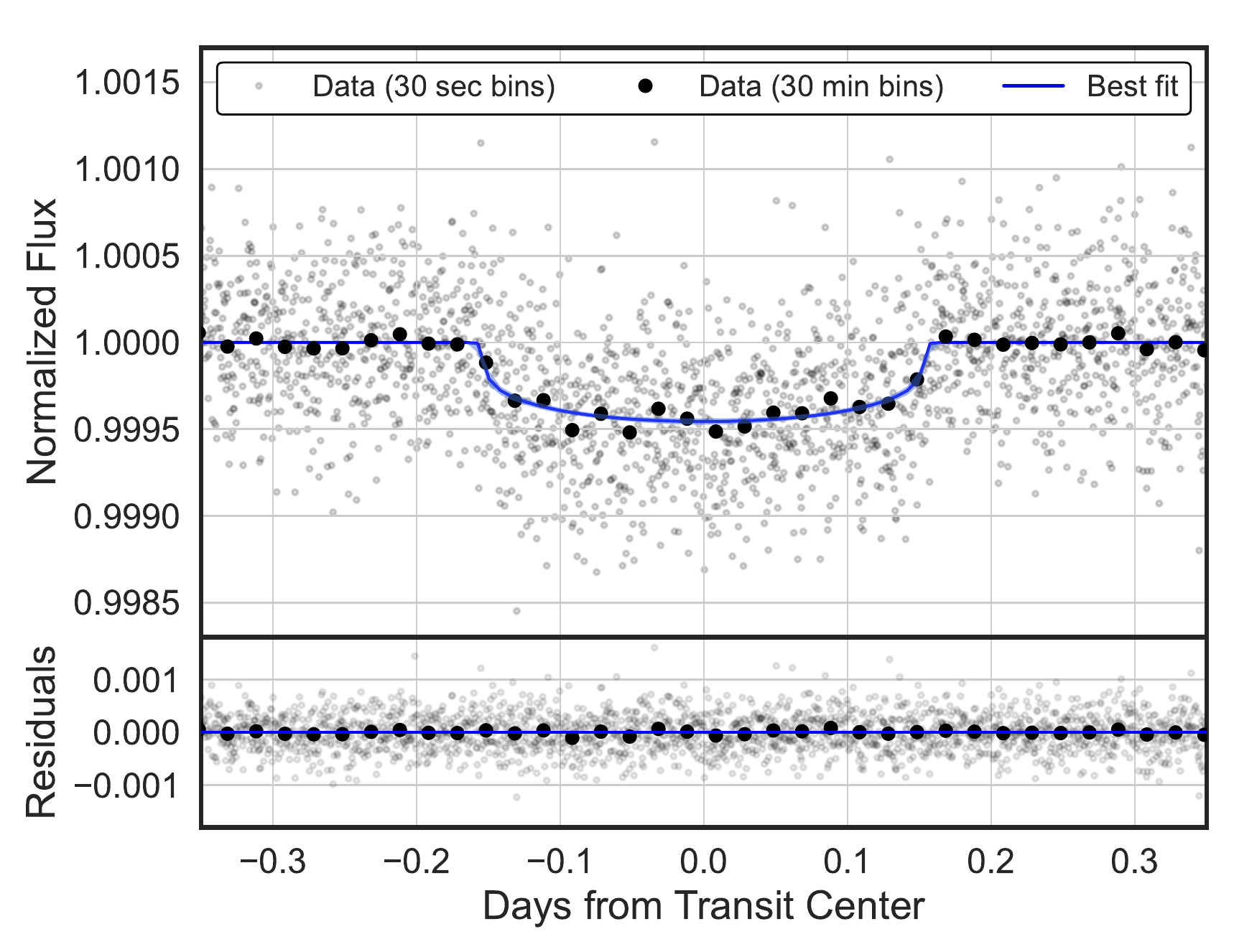}
    \caption{\emph{Top}: TESS lightcurve produced by the SPOC pipeline after flattening and outlier removal. Blue triangles show the times of transits used for fitting. The sectors are indicated at the top of the panel. Sector 4 was removed from the light curve analysis (see Section \ref{sec: TESS}). \emph{Bottom left}: SED fit of HD 56414. Blue points indicate broadband photometry measurements and the black curve indicates the best-fit stellar model with $T_{\rm eff} = 8500 \pm 150$ K. \emph{Bottom right}: Phase-folded TESS data centered on the transit with residual variability removed with a Gaussian Process model. The data is binned with bin widths of 30 seconds (gray points) and 30 minutes (black points). The blue line is the best-fit transit model.}
    \label{fig: data}
\end{figure*}

\section{Observations} \label{sec: observations}

\subsection{TESS Photometry} \label{sec: TESS}

HD 56414 (a.k.a. TOI-1228 or TIC 300038935) was observed by TESS for a total of 24 sectors between 2018 July 18 and 2021 June 24. During sectors 1--13, the target was observed with a cadence of 2 minutes. During sectors 28--36 and 38--39, the target was observed with cadences of both 2 minutes and 20 seconds. A transiting planet candidate, denoted TOI-1228.01, was identified in this data by the NASA Science Processing Operations Center \citep[SPOC;][]{jenkins2016} pipeline on 2019 August 26. The data and related analyses are available for download on the Mikulski Archive for Space Telescopes (MAST). The SPOC pipeline produces two light curves: a light curve produced using simple aperture photometry (SAP) and a light curve produced using simple aperture photometry plus a pre-search data conditioning step \citep[PDCSAP;][]{stumpe2014}. We utilized the latter for our analysis.

The PDCSAP process retains intrinsic stellar variability, which could influence interpretation of the data by introducing correlated noise. Specifically, stars in the temperature regime of HD 56414 are often $\delta$ Scuti variables, which can pulsate at periods under a day with amplitudes of a few ppt. While we found no evidence of periodic pulsations, we noticed some aperiodic variability of a few ppt that is likely residual background or instrumental noise. We removed long-term variability by fitting the data to a cubic basis spline with a knot distance of 1 day, linearly interpolating over the 7.6 hour-long transits to remove low frequency trends without altering the transit shape. Next, we removed short-term out-of-transit variability by clipping out the transits and fitting the light curve to a cubic basis spline with a knot distance of 0.1 day. We ensured that this process did not affect the morphologies of the transits by comparing each pre-flattening transit with its post-flattening counterpart. The transits of the planet candidate were identified using the orbital period ($P_{\rm orb}$) and time of first transit center ($T_0$) reported by the SPOC. Lastly, we removed outliers more than $5\sigma$ from the median flux. The resulting light curve is shown in Figure \ref{fig: data}.

Next, we inspected each transit and removed those with only partial coverage and those with poor data quality. Because our flattening procedure requires both pre- and post-transit baselines to properly normalize each transit, it had difficulty recovering the depths of partially detected transits. Consequently, partial transits that were significantly deeper than expected occurred near 1609.774 TBJD (where TBJD = BJD - 2457000) and  1638.824 TBJD due to gaps in data coverage. We also removed a transit that occurred at 1348.326 TBJD during sector 1, which took place following an incorrectly configured fine pointing calibration after a spacecraft momentum dump,\footnote{See sector 1 data release notes: \url{https://heasarc.gsfc.nasa.gov/docs/tess/cycle1_drn.html}} and all data from sector 4, during which a row of anomalously low pixels, caused by a smear correction applied to adjust for a saturated star located elsewhere on the CCD, passed directly through HD 56414. Our final dataset included 14 transits during 24 TESS sectors (see Figure \ref{fig: data}).

\subsection{SMARTS/CHIRON Spectroscopy}\label{sec: CHIRON}

We obtained reconnaissance spectra of HD 56414 on 2019 October 21, 2021 September 26, and 2022 January 13 using the CHIRON spectrograph on the 1.5 m SMARTS telescope, located at Cerro Tololo Inter-American Observatory (CTIO), Chile \citep{tokovinin2013}. The three spectra yielded signal-to-noise ratios per resolution element of 36.6, 71.5, and 117.5, respectively, at 550 nm. Because stellar properties can be difficult to extract from spectra of hot, rapidly-rotating stars, we primarily used these spectra to determine the line-of-sight rotational velocity of the star. To do so, we obtained a line profile from each spectrum via a least-squares deconvolution \citep{1997MNRAS.291..658D} between the observation and a synthetic non-rotating spectral template generated from the ALTAS-9 model atmospheres \citep{2003IAUS..210P.A20C}. We then modeled the resulting line profile via a line broadening kernel that incorporates the effects of rotational, instrumental, and macroturbulent broadening. This calculation yielded $v\sin{i_\star} = 59.4\pm3.0 \, {\rm km} \, {\rm s}^{-1}$. We also use the spectra to rule out false-positive-producing scenarios, which we discuss further in Section \ref{sec: validation}.

\subsection{SOAR/HRCam Speckle Imaging}

We obtained $I$-band Speckle interferometric observations of HD 56414 on 2019 March 17 with HRCam mounted on the 4.1 m Southern Astrophysical Research (SOAR) telescope, located at Perro Pach\'{o}n, Chile. These observations and their related analyses are outlined in \cite{ziegler2020, ziegler2021}, to which we direct the reader for more information. The observations achieved a contrast of 5.2 magnitudes at $1\arcsec$. The speckle auto-correlation function, which covers a field of view of $3\arcsec \times 3\arcsec$, and corresponding contrast curve are shown in Figure \ref{fig: vetting}. We use these observations to constrain the possibility of false-positive-producing scenarios, which we discuss further in Section \ref{sec: validation}.

\section{Data Analysis} \label{sec: analysis}

\subsection{Adopted Stellar Parameters} \label{sec: stellar}

Our adopted stellar parameters were obtained using a number of sources and methods (see Table 1). We adopted the stellar right ascension ($\alpha$), declination ($\delta$), proper motion in right ascension ($\mu_\alpha$), proper motion in declination ($\mu_\delta$), and parallax ($\pi$) from Gaia eDR3 \citep{gaia2021}.

Next, we used the broadband spectral energy distribution (SED) and Gaia eDR3 parallax of the star to determine the bolometric flux ($F_\mathrm{bol}$), effective temperature ($T_{\rm eff}$), and radius ($R_\star$) of the star \citep[following the procedures described in][]{stassun2016, stassun2017, stassun2018}. We obtained $G_{\rm BP}$ and $G_{\rm RP}$ from Gaia eDR3, our $B$ and $V$ magnitudes from APASS \citep{apass2016}, our $J$, $H$, and $K_s$ magnitudes from 2MASS \citep{2MASS2006}, and our $W1 - W4$ magnitudes from ALLWISE \citep{allwise2014}. We constrained interstellar extinction ($A_V$) using the analysis by \cite{kounkel2019} and stellar metallicity ([Fe/H]) based on the designation of HD 56414 as a metallic-line star by \cite{houk1975}. We fit the Kurucz stellar atmospheric models \citep{kurucz1993}, resulting in a reduced $\chi^2$ of 1.7 (see Figure \ref{fig: data}). We calculated the luminosity of the star ($L_\star$) directly using $F_{\rm bol}$ together with the Gaia distance, and stellar mass ($M_\star$) using the empirical relations presented in \cite{torres2010}. All of our derived properties are consistent with those quoted in version 8.2 of the TESS Input Catalog \citep{stassun2019}.

Lastly, we extracted the $v\sin{i_\star}$ and macroturbulent velocity ($v_{\rm mac}$) from the CHIRON spectra (Section \ref{sec: CHIRON}). Using this $v \sin{i_\star}$ and the $R_\star$ calculated above, we calculated $P_{\rm rot} / \sin{i}$ (where $P_{\rm rot}$ is the stellar rotation period). The rotation rate of the star is not informative of the stellar age because the star is hotter than the Kraft break. However, we assign an age of $420 \pm 140$~Myr based on its Theia 797 membership assigned in \cite{kounkel2019}, which identified stellar strings and calculated their ages using a combination of machine learning analysis and isochrone fitting with a typical uncertainty of 0.15 dex.

\begin{deluxetable}{lcc}[t!] \label{tab1}
\centering
\tabletypesize{\scriptsize}
\tablewidth{0pt}
\tablecaption{Adopted Stellar and Planet Parameters} 
\tablehead{\colhead{Parameter} & \colhead{Value} & \colhead{Source} }
\startdata
\multicolumn{3}{c}{Identifiers}\\
\hline
HD & 56414 & \\
TOI & 1228 & \\ 
TIC & 300038935 &   \\
2MASS & J07112249$-$6850006 & \\
Gaia DR2 & 5268547016621360768 &  \\
\hline
\multicolumn{3}{c}{Astrometry \& Position}\\
\hline
$\alpha$  & 07:11:22.48  & Gaia eDR3 \\
$\delta$  & -68:50:00.03 & Gaia eDR3 \\ 
$\mu_\alpha$ (mas\,yr$^{-1}$)& $-0.735\pm0.019$	& Gaia eDR3\\
$\mu_\delta$  (mas\,yr$^{-1}$) & $35.120\pm0.015$  & Gaia eDR3\\
$\pi$ (mas) & $3.735\pm0.013$ & \emph{Gaia} eDR3\\
Distance (pc) & $267.74\pm1.07$ & From parallax \\
\hline
\multicolumn{3}{c}{Photometry}\\
\hline
$G_{\rm BP}$ & $9.2908\pm0.0006$ & Gaia eDR3 \\
$G_{\rm RP}$ & $9.0508\pm0.0006$ & Gaia eDR3 \\
$B$ & $9.390\pm0.002$ & APASS \\
$V$ & $9.22\pm0.03$ & APASS \\
$J$ & $8.89\pm0.03$ & 2MASS \\
$H$ & $8.85\pm0.05$ & 2MASS \\
$K_s$ & $8.82\pm0.02$ & 2MASS \\
$W1$ & $8.78\pm0.02$ & ALLWISE \\
$W2$ & $8.80\pm0.02$ & ALLWISE \\
$W3$ & $8.82\pm0.02$ & ALLWISE \\
$W4$ & $ 9.1\pm0.2$ & ALLWISE \\
\hline
\multicolumn{3}{c}{Stellar Physical Properties}\\
\hline
$M_\star$ ($M_\odot$)& $1.89\pm0.11$& This paper \\
$R_\star$ ($R_\odot$)& $1.751\pm0.065$ & This paper \\
$T_{\rm eff}$ (K) & $8500\pm150$ & This paper \\
$L_\star$ ($L_\odot$) & $14.39\pm0.34$ & This paper \\
${\rm [Fe/H]}$ (dex) & $0.0_{-0.0}^{+0.3}$ & \cite{houk1975} \\
$A_V$ (mag) & $0.22\pm0.03$ & \cite{kounkel2019} \\
$v \sin{i_\star}$ (km\, s$^{-1}$) & $59.4\pm3.0$ & This paper\\
$v_{\rm mac}$ (km\, s$^{-1}$) & $3.2\pm1.0$ & This paper\\
$F_{\rm bol}$ (erg\,cm$^{-2}$\,s$^{-1}$) & $(6.44\pm0.15)\times10^{-9}$ & This paper \\
$P_{\rm rot} / \sin{i_\star}$ (days) & $1.49\pm0.09$ & This paper \\
Age (Myr) & $420\pm140$ & This paper  \\
\hline
\multicolumn{3}{c}{Best-Fit Planet Model Parameters}\\
\hline
$T_0$ (BJD) & $2458348.324\pm0.005$ & This paper \\
$P_{\rm orb}$ (days) & $29.04992\pm0.00021$ & This paper \\
$T_{\rm dur}$ (hours)$^*$ & $7.58\pm0.14$ & This paper \\
$R_{\rm p}$ ($R_\oplus$) & $3.71\pm0.20$ & This paper \\
$b$ & $0.34\pm0.19$ & This paper \\
$i$ ($^\circ$) & $89.3\pm0.4$ & This paper \\
$a$ (AU) & $0.229\pm0.004$ & This paper \\
$e$ & $< 0.68 \, (3\sigma)$ & This paper \\
$u_{\star,0}$ & $0.54\pm0.27$ & This paper \\
$u_{\star,1}$ & $0.19\pm0.37$ & This paper \\
$T_{\rm eq}$ (K)$^\dagger$ & $1133 \pm 30$ & This paper \\
ESM & $1.9 \pm 0.2$ & This paper \\
TSM$^\ddag$ & $30.0 \pm 1.7$ & This paper 
\enddata
\tablenotetext{*}{Calculated as the total transit duration, $T_{14}$.}
\tablenotetext{\dagger}{Calculated assuming zero Bond albedo and full day-night heat redistribution.}
\tablenotetext{\ddag}{Planet mass estimated using the mass--radius relation defined in \cite{chen2017} for Neptunian worlds.}
\end{deluxetable}

\subsection{Planet Model Fit}

We fit the TESS photometry to a transit model using \texttt{exoplanet} \citep{foreman-mackey2021}. The transit model was initialized with a quadratic limb darkening law (with coefficients $u_{\star, 0}$ and $u_{\star, 1}$) and the following priors: (1) Gaussian priors for $M_\star$, $R_\star$, $T_{\rm eff}$, $K_s$ magnitude, $T_0$, the logarithm of $P_{\rm orb}$, the logarithm of the transit depth, and the flux zero point of the light curve (2) a uniform prior on the transit impact parameter ($b$), and (3) a prior on eccentricity ($e$) based on \cite{kipping2013}. In addition, we included a Gaussian Process model with a simple harmonic oscillator kernel to account for residual variability in the light curve.\footnote{Although, we note that models with and without the Gaussian Process model did not yield significantly different results.} We ran a 10 walker ensemble with 20,000 steps and discarded the first 10,000 steps as burn-in. The best-fit phase-folded model is shown in Figure \ref{fig: data}.

As shown in Table 1, we fit for the following physical and orbital parameters: $T_0$, $P_{\rm orb}$, the transit duration $T_{\rm dur}$, the planet radius ($R_{\rm p}$), $b$, the orbital inclination ($i$), the orbital semi-major axis ($a$), $e$, $u_{\star, 0}$, and $u_{\star, 1}$. In addition, we estimated the planet equilibrium temperature assuming zero bond albedo and full day-night heat redistribution ($T_{\rm eq}$) and the planet emission spectroscopy and transmission spectroscopy metrics (ESM and TSM) assuming a circular orbit.\footnote{ESM and TSM are quantities defined by \cite{kempton2018} that indicate the suitability of the planet to emission and transmission spectroscopy. For TSM, we estimated the planet mass using the mass--radius relation defined in \cite{chen2017} for Neptunian worlds.} The best-fit model is a planet with $R_{\rm p} = 3.71 \pm 0.20\, R_\oplus$ on a $29.04992 \pm 0.00021$ day orbit with $e < 0.68$ ($3 \sigma$).

Based on the rapid rotational rate of HD 56414, we might expect to see an asymmetrical transit profile due to gravity darkening, which can be modelled to independently measure the true spin-orbit angle \citep[$\Psi$;][]{barnes2009}. To test this, we also fit the TESS light curve to a gravity-darkened model following \cite{hooton2022} in which we additionally fit for the stellar inclination ($i_\star$) and sky-projected stellar obliquity ($\lambda$). We fixed the stellar polar temperature, stellar equatorial radius, and $v\sin{i_\star}$ to the median values in Table 1, and fixed the gravity darkening coefficient $\beta$ according to \cite{claret2017}. In addition, we derived $P_{\rm rot}$, $\Psi$, and the stellar oblateness. The parameters shared between the gravity-darkened and non-gravity-darkened fits agreed well, while the gravity-darkened-specific model parameters were only weakly constrained. We constrained the stellar inclination to $i_\star > 19^\circ$ at the $3\sigma$ level, which rules out configurations with the rotation axis pointing towards the observer. This result is intuitive, as $P_{\rm rot}$ would need to be very short to explain the spectroscopically inferred $v\sin{i_\star}$ for small $i_\star$. Values of $i_\star$ between $45^\circ$ and $90^\circ$ were equally likely, resulting in $P_{\rm rot} = 1.33^{+0.23}_{-0.45}$ days. For $\lambda$ and $\Psi$, all angles are plausible, although polar orbits are slightly preferred. Better constraints on these angles may be obtainable with future transit observations. In addition, $\lambda$ could be obtained independently via the Rossiter-McLaughlin effect \citep[e.g.,][]{cameron2010}, although this would be difficult to accomplish given the shallow transit depth.

\section{Planet Validation} \label{sec: validation}

\begin{figure*}
    \centering
    \includegraphics[width=1.0\textwidth]{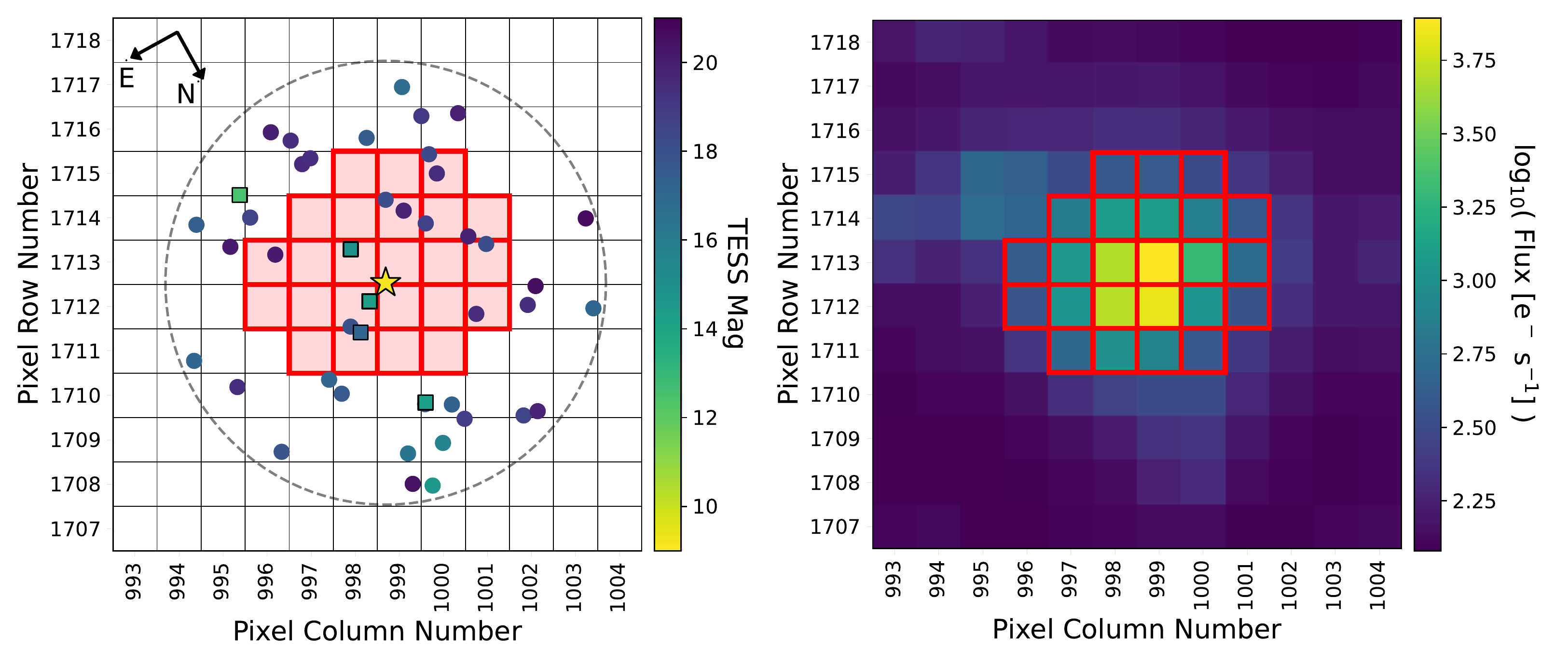}
    \includegraphics[width=0.45\textwidth]{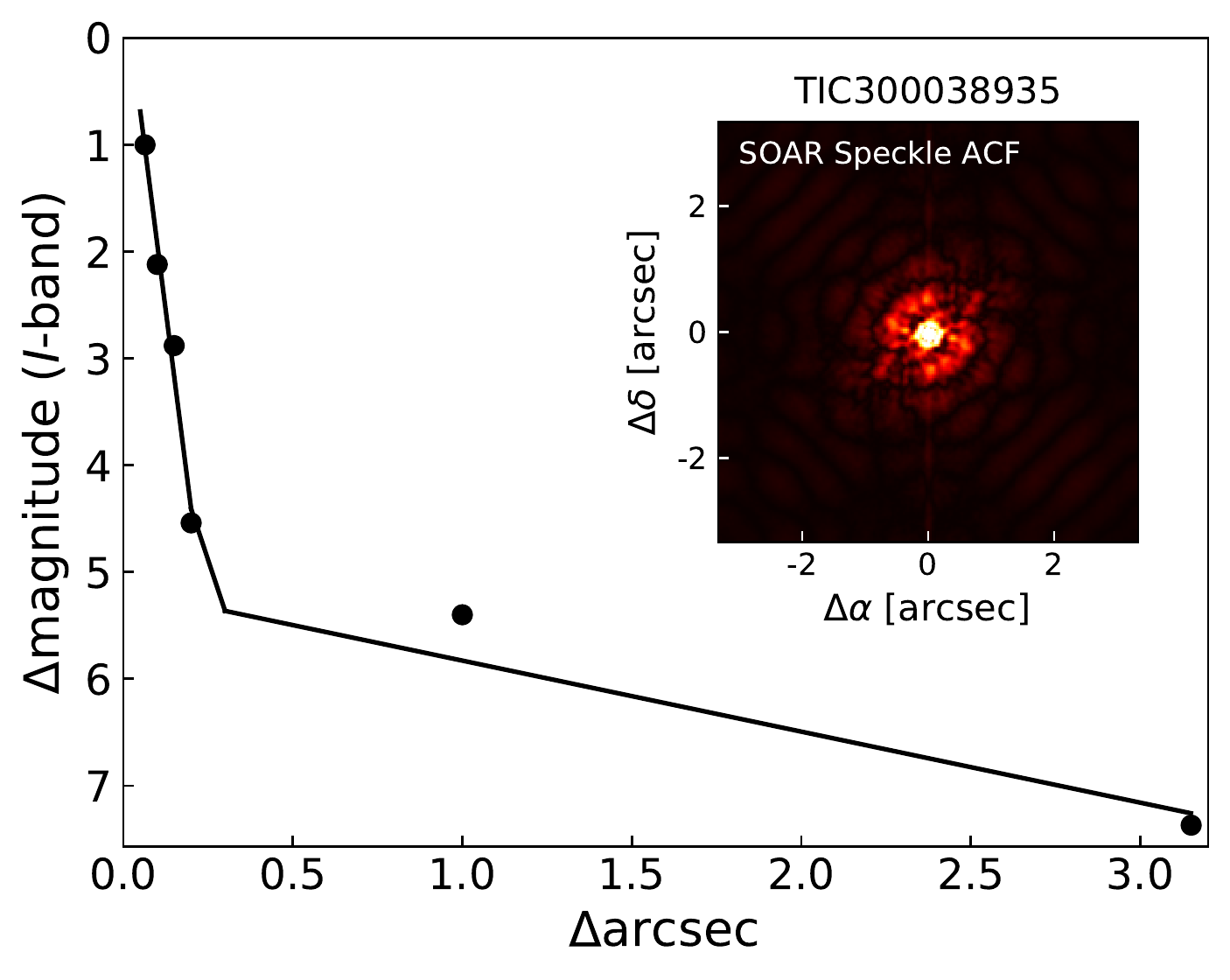}
    \includegraphics[width=0.49\textwidth]{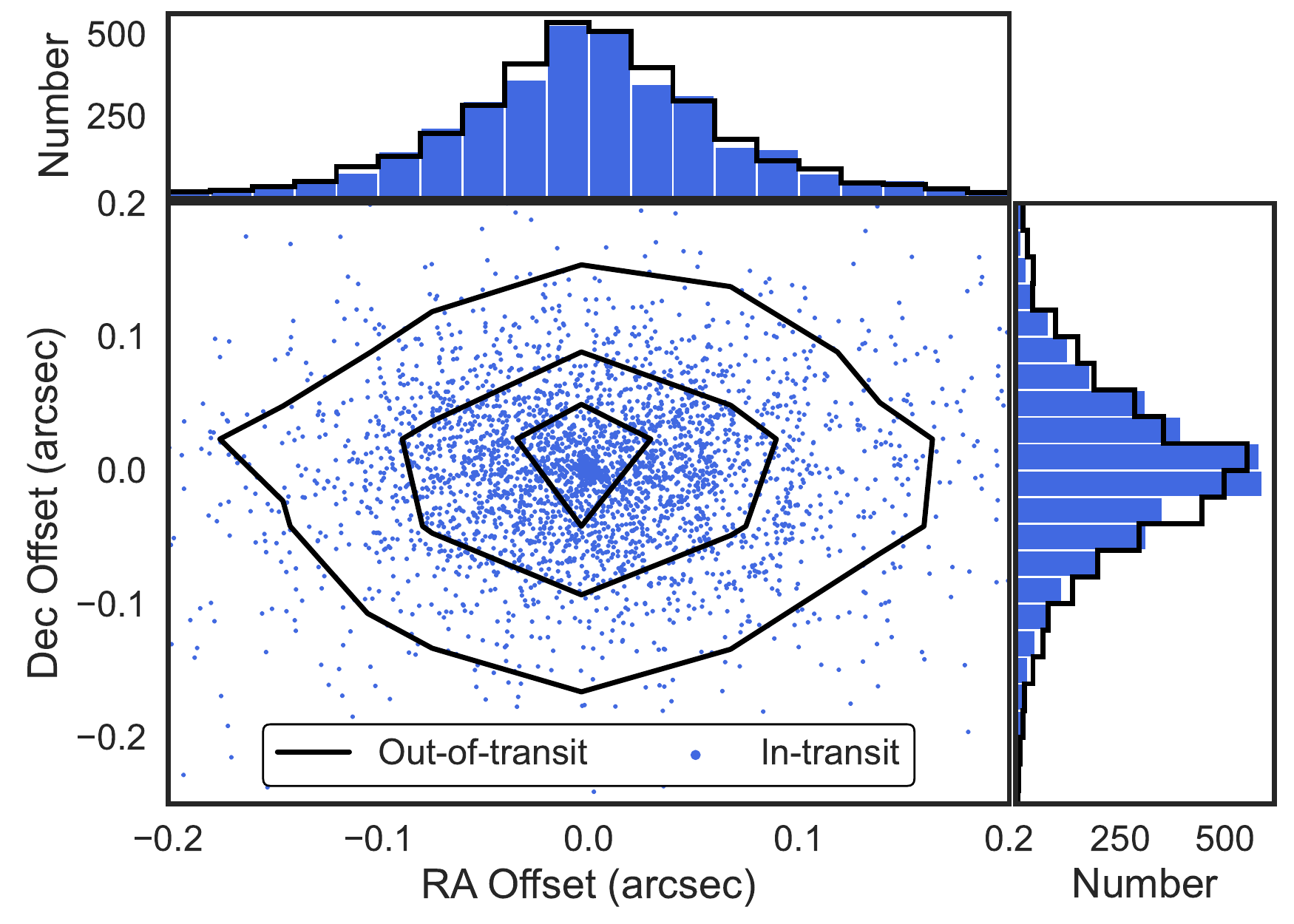}
    \caption{Visualizations of the vetting measures described in Section \ref{sec: validation}. \emph{Top-left}: Objects within $100\arcsec$ of HD 56414 (indicated by the yellow star). The SPOC aperture is shown in red, squares are nearby sources bright enough to be FP-inducing nearby eclisping binaries, and circles are nearby sources too faint to be FP-inducing nearby eclipsing binaries. \emph{Top-right}: Corresponding TESS image, with the same aperture overlaid. \emph{Bottom-left}: Speckle imaging image and contrast curve of HD 56414, which detects no companions within detection limits. \emph{Bottom-right}: Results of centroid offset analysis with \texttt{vetting}. HD 56414 is at the origin. Solid black lines are $1\sigma$, $2\sigma$, and $3\sigma$ contours formed from 3,629 randomly sampled out-of-transit data points. Blue points are centroids for each in-transit data point. The difference between the two resulting distributions is not statistically significant.}
    \label{fig: vetting}
\end{figure*}

\subsection{Follow-up Observations}

We utilized the ground-based follow-up observations described in Section \ref{sec: observations} to search for evidence of unresolved stellar-mass companions that would imply that TOI-1228.01 is a false positive (FP). We began by using our spectra to search for single-lined and double-lined spectroscopic binaries. The two high-S/N spectra were collected at 2483.8406 and 2592.6828 TBJD, which correspond to orbital phases of 0.09 and 0.84 using the orbital parameters in Table 1. These observations yielded RVs of $0.79 \pm 0.73$ km/s and $1.35 \pm 0.73$ km/s, respectively. At these orbital phases, a $0.1 \, M_\odot$ stellar companion on a circular 29-day orbit would produce a $\Delta \rm{RV} > 6$ km/s. These observations therefore rule out the possibility that the transit-like event is caused by a stellar-mass companion. This was corroborated by a search for double-lined spectroscopic binaries, which produced no evidence of excess stellar lines. 

We searched for evidence of longer period stellar companions using our speckle interferometric observations. Our observations rule out companion stars with $\Delta I < 1$ at orbital separations $> 0.1\arcsec$ and $\Delta I < 4.5$ at orbital separations $> 0.2\arcsec$, indicating that stellar-mass companions with brightnesses comparable to HD 56414 are unlikely to be present beyond 25 AU.

Nonetheless, these observations alone are not sufficient to validate TOI-1228.01 because some FP scenarios can evade detection by these methods. For instance, FPs can occur when the transit originates from a nearby source that is known, but is unresolved in the TESS data. To illustrate, Figure \ref{fig: vetting} shows that several additional stars lie near the aperture used to extract the light curve. We investigated these FP scenarios via the following statistical analyses.

\subsection{Centroid Offset Analysis}

FPs due to signals originating from nearby sources often cause offsets between in-transit and out-of-transit centroids, the photocenters of the target star in the TESS images \citep[e.g.,][]{bryson2013, twicken2018}. We searched for centroid offsets in the TESS data using the open-source tool \texttt{vetting} \citep{hedges2021}, which calculates a p-value for each TESS sector indicating whether or not an offset has been detected. For each sector in which a transit occurred, we obtained a p-value $> 0.05$, indicating no significant centroid offset. On average the in-transit centroids agree well with the out-of-transit centroids (see Figure \ref{fig: vetting}), providing strong evidence that the signal observed by TESS originates from HD 56414.

\subsection{Statistical Validation}

\texttt{TRICERATOPS} \citep{giacalone2021} is a Bayesian tool that can be used to validate TESS planet candidates by calculating a FP probability (FPP; the overall probability that the signal originates from something other than a planet transiting the target star) and a nearby FP probability (NFPP; the probability that the the signal originates from a known nearby source that is listed in the TIC), where the criteria for validation are ${\rm FPP} < 0.015$ and ${\rm NFPP} < 10^{-3}$. Folding in the speckle imaging data for an added constraint on the presence of unresolved FP-inducing stars, we ran \texttt{TRICERATOPS} 50 times and calculated the mean and standard deviation of the FPP and NFPP.\footnote{Multiple runs are necessary because the probability calculation is not deterministic.} We obtained ${\rm FPP} = 0.004 \pm 0.006$ and ${\rm NFPP} = (1.6 \pm 3.5) \times 10^{-7}$, which are sufficient for validation.\footnote{This FPP is dominated by the scenario in which an unresolved stellar companion hosts a transiting planet of a different size.} Hereafter, we refer to TOI-1228.01 as HD 56414 b.

\section{Discussion and Conclusions} \label{sec: discussion}

\begin{figure*}
    \centering
    \includegraphics[width=1.0\textwidth]{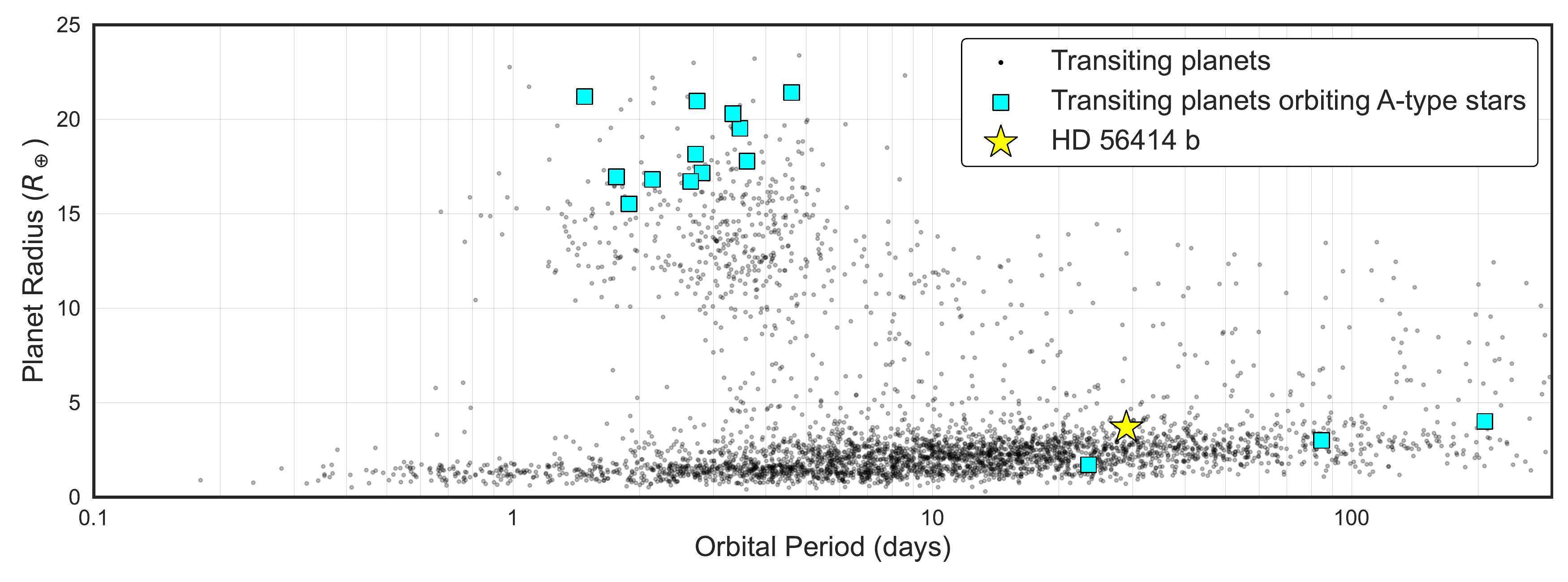}
    \includegraphics[width=1.0\textwidth]{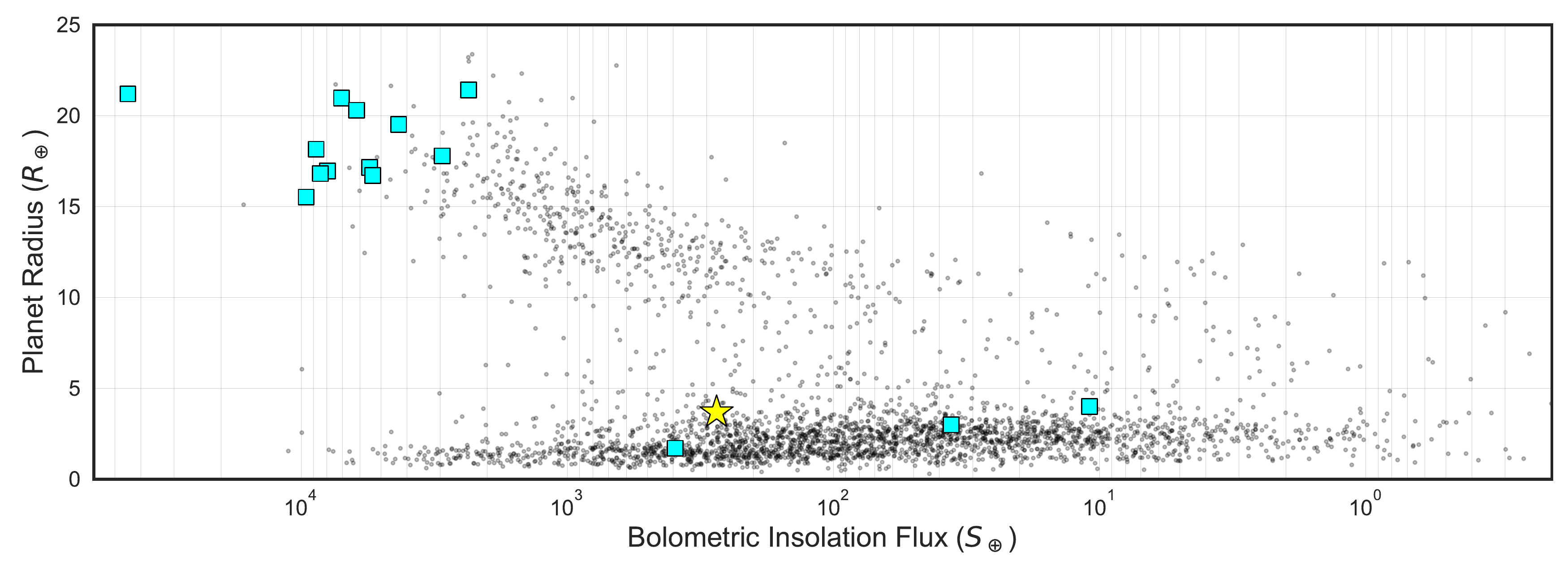}
    \includegraphics[width=1.0\textwidth]{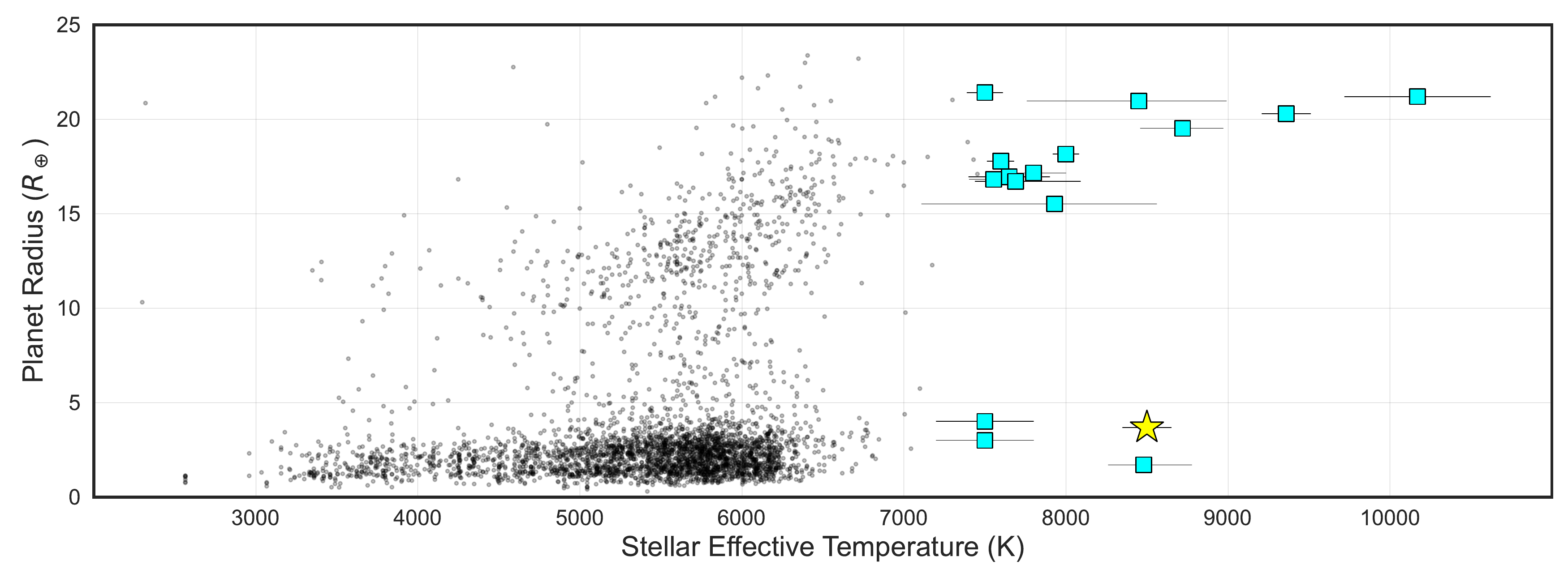}
    \caption{Confirmed transiting planets (cyan squares for A-type host stars; black dots for other host stars), and HD 56414 b (yellow star) in different parameter spaces. Error bars for planets orbiting A-type stars are shown, but are often smaller than the scale of the image. The three sub-Jovian cyan squares are (in order of increasing orbital period) Kepler-1115 b, Kepler-462 b, and Kepler-462 c. HD 56414 b is the second shortest period sub-Jovian planet known to exist around an A-type star and has one of the hottest host stars of any known sub-Jovian planet. Data acquired from the NASA Exoplanet Archive.}
    \label{fig: discussion}
\end{figure*}

HD 56414 b is a $3.71\pm0.20\, R_\oplus$ planet orbiting a $420 \pm 140$ Myr old A-type star on a 29 day orbital period. In Figure \ref{fig: discussion}, we compared HD 56414 b to all other confirmed transiting planets and confirmed transiting planets orbiting A-type stars. HD 56414 b is one of only 16 confirmed transiting planets orbiting an A-type star (which we define as any star with $7500 \, {\rm K} \leq T_{\rm eff} \leq 10500 \, {\rm K}$)\footnote{The upper limit is chosen to include KELT-9b \citep{gaudi2017}.} and one of only four confirmed sub-Jovian planets orbiting an A-type star. The other three sub-Jovian planets known to orbit A-type stars are Kepler-1115 b ($R_{\rm p} = 1.7 \, R_\oplus$, $ P_{\rm orb} = 23.6$ days), Kepler-462 b ($R_{\rm p} = 3.0 \, R_\oplus$, $P_{\rm orb} = 84.7$ days), and Kepler-462 c ($R_{\rm p} = 4.0 \, R_\oplus$, $P_{\rm orb} = 207.6$ days).\footnote{The spectral type of Kepler-462 is disputed, as some have estimated its $T_{\rm eff}$ to be under 7500 K \citep[e.g.,][]{berger2018}.} 

HD 56414 b is one of only 32 transiting planets orbiting a star with a well-constrained age (${\rm Age}/\sigma_{\rm Age} \ge 3$) that is $< 1$ Gyr (hereafter referred to as ``young'').\footnote{According to the NASA Exoplanet Archive, accessed 2022 May 18.} More specifically, this planet is one of 16 young transiting planets smaller than $4 \, R_\oplus$. Among these smaller planets, HD 56414 b is one of only 5 with $P_{\rm orb} > 20$ days. Further investigation of this system could therefore provide insight into the formation and early evolution of small, long-period planets. In addition, HD 56414 is the hottest and most massive young star with a known transiting planet, making it a valuable asset for understanding how these processes depend on stellar temperature and mass.

HD 56414 b shares several characteristics with Kepler-1115 b: the two planets orbit stars with similar effective temperatures ($T_{\rm eff} \sim 8500$ K) and have similar orbital periods, leading to comparable bolometric insolation fluxes ($S_{\rm bol}$). Figure \ref{fig: discussion} shows these two planets near what is conventionally referred to as the hot Neptune desert, a region of parameter space at short orbital periods (or, in this case, high $S_{\rm bol}$) that exhibits a dearth of planets with sizes of roughly $3-10 \, R_\oplus$ \citep{mazeh2016}. The fact that these are the two shortest period planets orbiting A-type stars, despite the bias that planets with shorter orbital periods are more likely to transit and are easier to detect, may indicate that sub-Jovian planets with $P_{\rm orb} < 20$ days are intrinsically rare around these hot stars.

\begin{figure}[t!]
    \centering
    \includegraphics[width=0.5\textwidth]{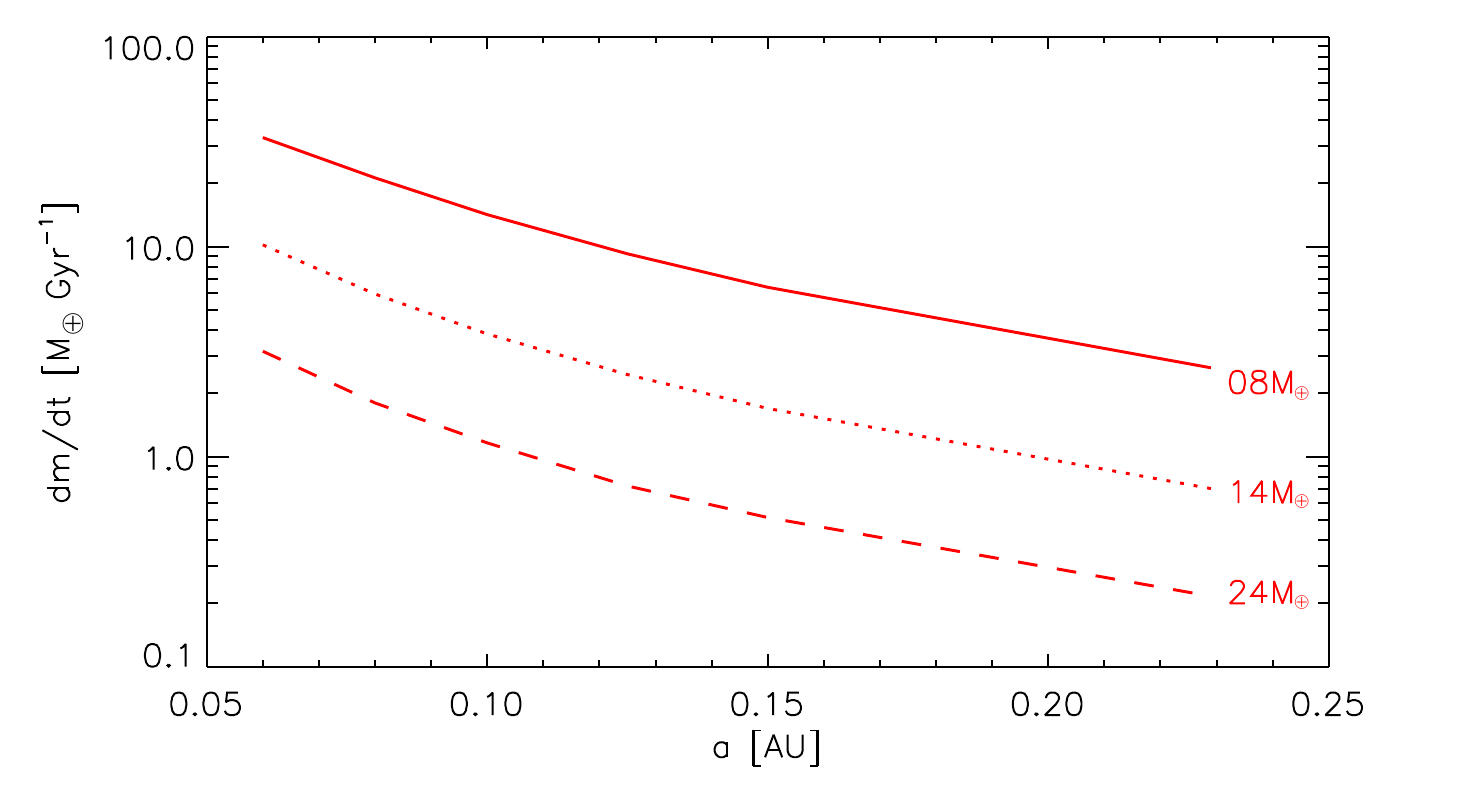}
    \caption{Simulated atmospheric mass loss rates as a function of orbital separation for planets with the same size as HD 56414 b orbiting a star identical to HD 56414. Each line represents a planet of a different mass. Assuming a core-to-envelope mass fraction of $\sim 10\%$, planets at $0.229$ AU with $M_{\rm p} = 8 \, M_\oplus$ can lose over a third of their atmospheric mass over the typical 1-Gyr lifetime of an A-type star. Closer in planets with $M_{\rm p} < 14 \, M_\oplus$ are susceptible to losing all of their atmospheres over this timescale.}
    \label{fig: photoevaporation}
\end{figure}

Some have proposed that the hot Neptune desert is a product of photoevaporation, the process by which planets are stripped of their atmospheres due to stellar emission of XUV ($\lambda < 912$ \r{A}) photons \citep[e.g.,][]{lopez2013, owen2013, owen2018}. \cite{munoz2019} showed that photoevaporation could be even more efficient at stripping planets of their atmospheres around A-type stars due to high stellar continuum emission beyond the Balmer limit ($\lambda < 3646$ \r{A}) in the near ultraviolet. To determine if a close-in Neptune-size planet could survive around such a hot star, we applied the same model for HD 56414 b. We used a PHOENIX LTE spectrum with $T_{\rm eff} = 8600 \, {\rm K}$, $\log{g} = 4.0$, and ${\rm [Fe/H]} = 0.0$ \citep{husser2013} and assumed an XUV flux of $3.6$ erg\,cm$^{-2}$\,s$^{-1}$ at 1 AU. This XUV flux is the same as that assumed in \cite{munoz2019} for KELT-9 b, which is probably similar to HD 56414 because both stars are in a temperature regime where it is unlikely for a stellar corona to be present. Nonetheless, we found that our results are not strongly sensitive to the choice of XUV flux because the mass loss rate is dominated by photons in the near ultraviolet. Because the mass of HD 56414 b is unknown, we estimated the range of possible masses of the planets using multiple mass--radius relations defined by other studies. Using the relation defined in \cite{chen2017}, we generated a distribution of masses centered just below $14 \, M_\oplus$ with $1\sigma$ lower and upper limits of $8 \, M_\oplus$ and $24 \, M_\oplus$, respectively. These results broadly agree with the corresponding relation defined in \cite{otegi2020}, which estimates a mass of $13.8 \pm 3.7 \, M_\oplus$. To account for this range of possible masses, we ran the model on planets with masses of $M_{\rm p} = 8 \, M_\oplus$, $14 \, M_\oplus$, and $24 \, M_\oplus$ on circular orbits with separations between 0.06 AU (corresponding to $P_{\rm orb} = 3.9$ days) and 0.229 AU (the best-fit semi-major axis). The resulting mass loss rates over a 1 Gyr timescale are shown in Figure \ref{fig: photoevaporation}.

If we assume a core-to-envelope mass fraction of $\sim 10 \%$, our results suggest that HD 56414 b could lose an over a third of its atmospheric mass over 1 Gyr if $M_{\rm p} = 8 \, M_\oplus$. In addition, closer in planets with $M_{\rm p} < 14 \, M_\oplus$ are susceptible to losing nearly all of their atmospheric mass. These results imply that, unless they typically more massive than $14 \, M_\oplus$, Neptune-size planets within 0.1 AU of A-type stars are rare due to rapid stripping of their atmospheres. Consequently, if we assume that Neptune-size planets can acquire these close-in orbits via in situ formation or disk migration \citep[e.g.,][]{lee2017}, we might expect to see a pile-up of remnant Neptunian cores at short orbital periods \citep{lecavelier2004}.

It may be possible to deepen our understanding of HD 56414 b by obtaining additional observations. As discussed in Section \ref{sec: analysis}, the line-of-sight stellar obliquity can be revealed with additional transit or Rossiter-McLaughlin observations. Also, if the planet is actively undergoing atmospheric escape, it may be possible to detect at the Lyman-$\alpha$ \citep[e.g.,][]{vidal-madjar2003}, H$\alpha$ \citep[e.g.,][]{yan2018}, or HeI 10830 \AA \, lines \citep[e.g.,][]{spake2018}. It is also possible to study this planet via emission and transmission spectroscopy with the James Webb Space Telescope (JWST) or other facilities. We quantified these prospects by calculating the ESM and TSM for the planet, which \cite{kempton2018} recommend be above 7.5 and 90, respectively, for planets of similar sizes.\footnote{Technically, \cite{kempton2018} only calculated a recommended minimum ESM for planets with $R_{\rm p} < 1.5 \, R_\oplus$, but we quote that number in the absence of a recommendation for larger planets.} We calculated relatively low values of $1.9 \pm 0.2$ and $30.0 \pm 1.4$, respectively, owing to the small planet-to-star radius ratio of HD 56414 b. Nonetheless, the brightness of HD 56414 and its proximity to the southern JWST continuous viewing zone makes it a significantly better target than Kepler-1115 ($V = 11.98$, $K_s = 11.51$) and Kepler-462 ($V = 11.63$, $K_s = 10.85$). Given that HD 56414 b occupies a sparsely populated region of parameter space, it makes a compelling target for a follow-up  campaign to characterize its atmosphere.

\begin{acknowledgements}

We thank Eugene Chiang for the constructive conversations that led to the analyses in this paper. This work is supported by NASA FINESST award 80NSSC20K1549 (FI: Giacalone, PI: Dressing). CDD also acknowledges supported provided by the David and Lucile Packard Foundation via Grant 2019-69648. SQ acknowledges support from the TESS GI Program under award 80NSSC21K1056. GZ thanks the support of the ARC DECRA program DE210101893. Based in part on observations obtained at the Southern Astrophysical Research (SOAR) telescope, which is a joint project of the Minist\'{e}rio da Ci\^{e}ncia, Tecnologia e Inova\c{c}\~{o}es (MCTI/LNA) do Brasil, the US National Science Foundation’s NOIRLab, the University of North Carolina at Chapel Hill (UNC), and Michigan State University (MSU). We acknowledge the use of public TESS data from pipelines at the TESS Science Office and at the TESS Science Processing Operations Center. Resources supporting this work were provided by the NASA High-End Computing (HEC) Program through the NASA Advanced Supercomputing (NAS) Division at Ames Research Center for the production of the SPOC data products. This research has made use of the NASA Exoplanet Archive, which is operated by the California Institute of Technology, under contract with the National Aeronautics and Space Administration under the Exoplanet Exploration Program.

\end{acknowledgements}

\software{\texttt{exoplanet} \citep{foreman-mackey2021}, \texttt{vetting} \citep{hedges2021}, \texttt{TRICERATOPS} \citep{giacalone2021}}

\bibliography{references.bib}

\end{document}